\documentclass{PoS}

\usepackage{graphicx}
\usepackage{amsmath}
\usepackage{amsfonts}
\usepackage{wrapfig}
\usepackage{mathrsfs}
\usepackage{amssymb}
\usepackage{psfrag}
\usepackage{epsfig}

\newlength{\intwidth}

\newcommand{\EQ}{\begin{equation}}
\newcommand{\EN}{\end{equation}}
\newcommand{\bea}{\begin{eqnarray}}
\newcommand{\eea}{\end{eqnarray}}

\newcommand{\Z}{\mathbb{Z}}

\newcommand{\R}{\mathscr{R}}
\newcommand{\U}{\mathscr{U}}
\newcommand{\bra}{\langle}
\newcommand{\ket}{\rangle}
\newcommand{\avg}[1]{\langle \hspace{0.2em} #1 \hspace{0.2em} \rangle}

\newcommand{\sun}{\mathop{\rm SU}(N)}

\newcommand{\eq}{\begin{equation}}
\newcommand{\en}{\end{equation}}
\newcommand{\ea}{\end{eqnarray}}

\newcommand{\link}[1]{\bra #1\ket}

\title{On the ratio of string tensions in the 3D ${\mathbb Z}_4$ lattice gauge theory}

\ShortTitle{On the ratio of string tensions in the 3D ${\mathbb Z}_4$ lattice gauge theory}
\author{M.~Caselle, P.~Giudice, F.~Gliozzi, \speaker{P.~Grinza} and S. Lottini \\
       Dipartimento di Fisica Teorica, Universit\`a di Torino and\\
       INFN, Sezione di Torino \\
       via P. Giuria 1, I-10125 Torino, Italy \\
        E-mail: \email{caselle,giudice,gliozzi,grinza,lottini@to.infn.it}}

\abstract{It was recently pointed out that simple scaling properties of Polyakov correlation functions of gauge systems in the confining phase suggest that 
the ratios of k-string tensions in the low temperature region is constant 
up to terms of order $T^3$. Here we argue that, at least in a 
three-dimensional ${\mathbb Z}_4$ gauge model, the above ratios 
are constant in the whole confining phase. This result is obtained by 
combining numerical experiments with known exact results on the mass spectrum 
of an integrable two-dimensional spin model describing the infrared behaviour of 
the gauge system near the deconfining transition.}

\FullConference{The XXV International Symposium on Lattice Field Theory\\
		 July 30 - August 4 2007\\
		 Regensburg, Germany}

\begin{document}
\renewcommand{\thefootnote}{\arabic{footnote}}
\setcounter{footnote}{0}
\section{Introduction}
% pstoedit -f fig eps.eps > eps.fig
It was recently pointed out that in most confining  gauge theories, besides the fundamental string (of tension $\sigma$) which is formed between a pair of static sources in the fundamental representation $f$, there is the freedom of taking the sources in any representation $\R$. 
If, for instance, the gauge group is $\sun$ there are infinitely many 
irreducible representations at our disposal. However, as the sources are
pulled apart, no matter what
representation is chosen, the asymptotically stable string tension 
$\sigma_{\R}$  depends only 
on the $N-$ality $k$ of $\R$.
%, i.e. on the number (modulo $N$) of copies of 
%the fundamental representation needed to build $\R$ by tensor product, 
%because
%all representations with the same $k$ can be transformed into each other
%by the emission of a proper number of soft gluons. 
As a consequence the
heavier strings decay into the string of smallest string tension $\sigma_k$. 
The corresponding string is referred to as a k-string. 
This kind of confining object can be defined whenever the gauge group admits more than one non trivial irreducible representation.
%Much work has been done in the study of k-string tensions in the continuum \cite{ds}--\cite{Armoni:2006ri} as well as
%on the lattice \cite{lt1}--\cite{ddpv}. 

In a previous work \cite{confanomaly}, some of us have proposed an expression for the low temperature asymptotic expansion for these string tensions. An interesting consequence of such an expansion is that their ratios are expected to be constant up to  $T^3$ terms.  
The low temperature data presented in support of this expectation were 
taken from Monte Carlo simulations on a particular system, namely a 
(2+1)-dimensional $\Z_4$ gauge model. 

The main conjecture we want to verify in this work is that 
$\sigma_k(T)/\sigma(T)$, at least in that $\Z_4$ gauge system, is in fact 
independent of the temperature in the \emph{whole} of the confining regime. 
To check this idea we used the fact that the Svetitsky-Yaffe (SY) conjecture 
\cite{sy} allows to reformulate the 
system in a totally different perspective, based on a two-dimensional 
integrable theory. 

It turns out that the deconfinement transition of the 3D 
$\Z_4$ gauge model is second order and, according to the SY conjecture,
  belongs to the same universality class of the 2D symmetric 
Ashkin-Teller (AT) model. 

The two-dimensional AT model can
be seen in the continuum limit as a bosonic conformal field theory plus a 
massive perturbation  driving the system away from the critical line 
(i.~e.~a sine-Gordon theory). Thus, a map between the AT critical line 
and the sine-Gordon phase space is provided.
This theory is integrable, and the masses of its lightest physical states 
(first soliton and first breather mode, of masses $M$ and $M_1$) 
correspond to the tensions
$\sigma(T)$ and $\sigma_2(T)$ near $T_c$, whose ratio, in this context, 
can be analytically 
evaluated and turns out to be
\eq
\lim_{T\to T_c}\frac{\sigma_2(T)}{\sigma(T)}=
\frac{M_1}{M}=2\sin\frac\pi2(2\nu-1)\;\;,
\label{main}
\en
where $\nu$ is the thermal exponent.

\subsection{The (2+1)d $\Z_4$ gauge model and its dual reformulation}
\label{dualref}

The most general form of $\Z_4$ lattice gauge model admits two independent coupling constants, with partition function
\EQ
	\Z(\beta_f,\beta_{ff}) = \prod_l \sum_{\xi_l = \pm1, \pm i} e^{\sum_p (\beta_f \U_p + \beta_{ff} \U_p^2/2 + \mathrm{c.c.})} \; ,
\EN
in which the gauge field $U_l$ on the links on a cubic lattice is valued among the fourth roots of the identity and the sum in the exponent is taken over the elementary plaquettes of the lattice.
Such a theory can be reformulated as two coupled $\Z_2$ gauge systems (see \cite{confanomaly} for details):
\EQ
	\Z(\beta_f,\beta_{ff}) = \prod_l \sum_{\{ U_l=\pm 1, V_l = \pm 1 \}} e^{\sum_p [ \beta_f(U_p+V_p) + \beta_{ff} U_p V_p ] } \;\;\;, 
(U_p=\prod_{l\in p}U_l
\;\; ; V_p=\prod_{l\in p}V_l\;).
\EN

From the data in \cite{confanomaly}, obtained by means of 
finite-temperature measurements of  Polyakov-Polyakov correlation functions, and particularly 
from those referring to the point $P$ identified by $(\alpha,\beta)=(0.050, 0.207)$, the string tensions $\sigma$ and $\sigma_2$ can be evaluated in the $T \to 0$ 
limit as temperature-independent quantities:
\bea
	\sigma\,a^2  =    0.02085(10) \nonumber \; , \ \ \ \ \ 
	\sigma_2\,a^2  =  0.03356(22)  \; ,
\ea
where $a$ is the lattice spacing.
Their ratio, which has been argued to equate the central charge of the CFT related to the 2-string, is then given by
\EQ
	\frac{\sigma_2}{\sigma} = 1.610(13) \; .
\label{ratiosigma}
\EN

% FINE STEO
% INIZIO PAOLO

\section{The Svetitsky-Yaffe conjecture and the Sine-Gordon model}
%\subsection{The scaling field theory}
The mapping induced by the Svetitsky-Yaffe conjecture leads to a substantial simplification in the study 
of the critical properties of the deconfining transition, allowing to study it as a standard symmetry-breaking 
transition which takes place in a spin model. In the present case we 
deal with the symmetric Ashkin-Teller model in two-dimensions.

The action for this model is given by:
\EQ
	\label{eq:2dataction}
	{\cal{S}}_{AT} = - \sum_{\link{xy}} [J(\sigma^1_x\sigma^1_y+\sigma^2_x\sigma^2_y)+J_4(\sigma^1_x\sigma^1_y\sigma^2_x\sigma^2_y)] \; .
\EN

Such a model has been extensively studied in the past, and a number of exact results have been derived 
\cite{baxter}. It is useful to note that it can be seen as a perturbation of the Gaussian model, and in such a bosonic language the thermal perturbation can be written as $\cos \beta \varphi$, where $\beta$ is a marginal parameter equivalent to $J_4$. Hence we are left with
\bea
{\mathcal A_{\textrm{\tiny AT}}} = \int d^2 x \, \left( \frac12 \partial_\mu \varphi \partial^\mu \varphi \, - \, \tau \,  \cos \beta \varphi
\right) \; ,
\eea
which is the action of the Sine-Gordon model. 
 Furthermore, since the confined phase of the gauge theory is mapped in the high-T phase of the Ashkin-Teller, we will only consider the case $\tau>0$. 

Such a QFT is of particular interest because it is integrable, and this is 
the main reason for rewriting the action of 
the model near the critical point in a bosonic form.
%\footnote{It is always possible to \emph{fermionise} the 
%action of the Sine-Gordon model in order to obtain an integrable fermionic theory with the same scattering matrix, namely the massive Thirring model.}
Integrability means that an infinite number of integrals of motion exists. The main consequence in (1+1) dimensions is the fact that the scattering theory is
very constrained, because the $S$-matrix is factorised in products of two-body interactions, and inelastic processes are forbidden. It follows that the latter can be computed exactly together with the mass spectrum.

\subsection{Operator correspondence, mass spectrum and correlation functions}
\label{sub-operator}

%The last ingredient we need before exploiting the map to the Sine-Gordon model at its best, is the correspondence between the Polyakov loops in higher representation and the operators of the Ashkin-Teller model.
%The original paper by Svetitsky-Yaffe already provides the correspondence between the Polyakov loop in the fundamental representation and the spin operator.
We already know, from the Svetitsky-Yaffe original work, that the Polyakov loop 
in the fundamental representation corresponds to the spin operator. 
Then, following the same reasoning used in \cite{kdk}, it is possible to deduce that the Polyakov loop in the double fundamental representation is related to the so-called \emph{polarisation} operator ${\mathcal P} = \sigma^1  \sigma^2$,
where $\sigma^1$ and $\sigma^2$ are the spin variables defined in 
(\ref{eq:2dataction}). Its bosonic form and the corresponding  anomalous 
dimensions are given by
\bea
{\mathcal P} = \sin \frac{\beta}{2} 
\varphi, \ \ \ \ \ X_{\mathcal P} = \frac{\beta^2}{8 \pi} \; ;
\eea
we also notice that $\langle {\mathcal P} \rangle =0 $ in the 
high-T phase of the model.

\underline{Sine-Gordon mass spectrum \cite{refsuSGsmatrix}:} The exact 
knowledge of the S-matrix allows to access to the exact mass spectrum of the theory. Without entering the details, the spectrum of the SG model is given by a soliton/anti-soliton doublet of fundamental particles of mass $M$, and a number of soliton/anti-soliton bound states, called breathers $B_n$, whose number is a function of $\beta^2$.
By defining the coupling constant $\xi$ in the following way
\bea
\xi = \frac{\pi \, \beta^2}{8 \pi - \beta^2} \; ,
\eea
we have that for $\xi \ge \pi$, i.e. $\beta^2 \ge 4 \pi$, no bound states 
are present and hence the spectrum is given by the soliton/anti-soliton 
doublet only (repulsive regime). For $\xi < \pi$, i.e. $\beta^2 < 4 \pi$, we are in the attractive regime 
and the breathers $B_n$ appear as simple poles of the S-matrix (see for example \cite{refsuSGsmatrix}). 
The next step is to associate particle states to operators in the 
high temperature phase. It has been done in \cite{aldopaolo} by 
taking into account their properties of symmetry and locality. 
The consequence is that the spin operator is naturally associated to the mass of the soliton, and the polarisation operator is associated to the mass $M_1$ of the breather $B_1$. 
Hence, following the Svetitsky-Yaffe conjecture, the ratio of string tensions in the confining phase near the transition is given by
\bea
\frac{M_1}{M} \ = \ 2 \, \sin \frac{\xi }{2} \; .
\eea
This result, being a dimensionless ratio, is expected  to be universal in 
the limit $\tau \to 0$. This fact can be explicitly seen by expressing the 
coupling 
$\xi$ in terms of some critical exponent.
It is possible to work out the following relation between $\xi$ and the thermal critical exponent $\nu$
\bea
\label{eq:ratio_nu}
	\xi \    = \pi \, (2\nu-1) \ \ \ \   \to \ \ \ \ 
\frac{M_1}{M} \ = \  2 \, \sin \frac{\pi}{2}  \, (2\nu - 1) \;.
\eea

\underline{Correlators at large distance:} The previous analysis of the mass spectrum allows to compute the leading behaviour of the correlators $\langle \sigma \sigma \rangle $ and $\langle  {\mathcal P}   {\mathcal P}  \rangle$ at large distance by means of their spectral expansion over form factors (the interested reader can refer to \cite{smiyuzam,z3potts} for the details).

 The analysis of the previous section allows immediately to write down the leading term for $\langle \sigma \sigma \rangle $ and $\langle  {\mathcal P}   {\mathcal P} \rangle$ correlators in the high-T phase of the theory, up to an inessential proportionality constant
\bea 
\langle \sigma (x) \sigma (0) \rangle & \sim  & K_0 (M |x|), \ \ \ \ \ |x| \to \infty \nonumber \; ; \\
\langle  {\mathcal P} (x)   {\mathcal P} (0) \rangle & \sim  & K_0 (M_1 |x|), \ \ \ \ \ |x| \to \infty \; ,
\eea
where $K_0$ denotes the modified Bessel function of order zero, 
and $M$, $M_1$ are the masses of the soliton and the first breather 
respectively.  
%\vskip0.7cm

%\end{enumerate}

\subsection{Baryon vertices and mass spectrum}
\label{baryon}
%The general principle invoked in \cite{confanomaly} to derive formula (\ref{tension}) is simply that in a d-dimensional gauge theory whatever 
%correlation function made with Polyakov loops in the 
%\underline{fundamental} representation should be described, 
%at sufficiently low temperature and in the IR limit, by a two-dimensional 
%conformal field theory with central charge $c=d-2$.
%    
%A simple consequence of this general principle is that, as long as the 
%temperature is far from 
%the critical one, the shape of the world-sheet spanned by the baryon 
%vertices should be temperature independent; this ensures that the 
%baryon static potential has the expected asymptotic form \cite{confanomaly}.
% 
As noticed in \cite{confanomaly}, the balance of the string tensions 
for a given vertex gives the following expression for the angles 
at the center of the junction of three arbitrary k-strings
\bea
\label{angoli}
\cos \, \theta_i = \frac{\sigma^2_j (T) + \sigma^2_k (T) - \sigma^2_i (T)}{2\, \sigma_j (T)\sigma_k (T)}, \ \ \ \ \textrm{and cyclic permutations of 
the indices.}
\eea
The rigidity of the geometry of the vertex is then ensured by requiring that such angles are kept fixed when the temperature varies. 
As a consequence, all the string tension ratios are constant up to a given 
order in $T$, namely as far as the effective string picture is valid. 
%In other words, the previous geometrical construction is likely to 
%break down when the system approaches the deconfining temperature, 
%as the string begins to fluctuate wildly.

A similar  picture emerges when studying the gauge system near the 
deconfining transition. The scattering theory describing the system in such a case
can exhibit bound states whose mass $m_b$ is given by the following relation
% \cite{revgius}
\bea
m_b^2 = m_1^2 + m_2^2 + 2 m_1 m_2 \cos  u_{12}^b, \ \ \ \textrm{triangle of masses}
\label{triangle}
\eea 
where $\theta = i \, u_{12}^b$ is the purely imaginary value of the 
rapidity corresponding to the creation of the particle $m_b$, and $m_1$, $m_2$ are the masses of
the initial state.

In the present case the 
 process of coalescence of two fundamental strings into a 2-string 
corresponds to the scattering of a soliton/anti-soliton pair creating 
the bound state $B_1$. 
For such a process we know that $u_{\mathrm{S \bar S}}^{\mathrm{B}_1}=\pi-\xi$ which, once inserted in (\ref{triangle}),  gives
\bea
M^2_1 = 2 M^2 (1- \cos \xi )  \ \ \ \  \to  \ \ \ \   \frac{M_1}{M} = 
2 \sin \frac{\xi}{2}
\eea
which is nothing but the mass formula used in the previous Section.

\section{Monte Carlo setting and procedure}

\subsection{Mass ratio by correlators}

As introduced in Subsection~\ref{sub-operator}, 
we can determine the ratio $M_1/M$
using the large distance asymptotic behaviour of correlators; actually, 
exploiting the Svetitsky-Yaffe conjecture, we measured the 
Polyakov-Polyakov correlators $G_{\R}(R)$ of the (2+1)d $\mathbb{Z}_4$ gauge theory:
\eq
	G_{\R}(R)=\bra P_\R(0) P_\R^\dagger(R) \ket.
\en
In Section~\ref{dualref} we have explained we can study this theory
by means of simulations on the AT model and in~\cite{Giudice:2006hw}
the measurement of Polyakov-Polyakov correlators in both the fundamental
and double fundamental representations, $G(R)_{f}$ and $G(R)_{ff}$,
is described in detail.

%\FIGURE{
%\centering
%\includegraphics[angle=0, width=6cm]{plot-corr-f.eps}
%\caption{Polyakov-Polyakov correlator in the fundamental representation.}
%\label{fig:corr_f}
%}
We have taken $10^6$ measures on the $64^2 \times 7$ lattice in the phase
space point $P$; $N_\tau = 7$ is chosen because it is the lowest possible
value above the deconfinement transition. Simulations have been done for 
each value of $R$ in the range $[15 \div 44]$.
These data are fitted using an expansion of the $K_0(mR)$ Bessel function,
truncated to first two terms,
%\eq
%G(R)=cost \times \frac{e^{-mR}}{\sqrt{mR}} \left[ 1- \frac{1}{8 m R} \right]
%+ \mbox{``echo terms''},
%\en 
in a range $[R_{min},R_{max}]$, where $R_{max}=44$; we have verified the 
results are stable when $R_{min}$ varies in the range $[22 \div 33]$.
Therefore, it is possible to determine the two masses:
\bea
a\,M_{ff}  =  0.0698(15) \quad  (\chi^2/\mbox{d.o.f.} \approx 1.3) 
\nonumber \; , \ \ \ \ \ \ 
a\,M_f     =  0.0433(8)  \quad  (\chi^2/\mbox{d.o.f.} \approx  1.2) \; ,
\eea
from which we can determine the ratio:
\eq
\sigma_2(T\sim T_c)/\sigma(T\sim T_c)=M_{ff}/M_f =1.612(46)\,.
\label{mffovermf}
\en
This result, obtained near the critical temperature, is compatible 
with the zero-temperature value (\ref{ratiosigma}), providing
a strong evidence for our conjecture.

\subsection{Estimating $\sigma_2/\sigma$ through the thermal  exponent $\nu$ with finite-size scaling}

To use the formula for the mass ratio, Eq.~(\ref{eq:ratio_nu}), we need a quite precise estimate for the thermal critical exponent $\nu$ in the phase space point $P$. It can be obtained by means 
of a finite-size scaling analysis of the plaquette operator or some related observable that we denote with $\avg{\square}_L $, where $L$ is the spacial size of the lattice. Let us notice that in the SY context the plaquette operator 
is mapped into a combination of the unity and the energy operator of the 
corresponding CFT \cite{Gliozzi:1997yc}.

In order to exploit the computational advantages of the dual 
transcription of the gauge model, it is convenient to evaluate directly 
the internal energy of the 3D AT model defined by
\eq
\avg{\square}_L \equiv -\frac{1}{3L^2L_t} \avg{S_{AT}}\;\;.
\en
We decided however to use the corresponding (density of) susceptivity
\EQ
	\avg{\chi}_L \equiv \avg{ ( \square - \avg{\square}_L )^2 }_L \; ,
\EN
whose power-law to compare with has the form
\EQ
	\label{eq:suscettiva}
	\avg{\chi}_L = b' \cdot L^{\frac{2}{\nu}-d} \; ,
\EN
with the advantage that no constant additive terms are present, 
which could largely spoil the stability of the numerical results.

At the practical level, the system at the coupling $P$ turns out to be critical 
for a temperature $T_c$ such that $6 < \frac{1}{T_c} < 7$, hence, 
having to work with integer inverse temperatures, it is not possible 
to avoid some approximate method. In particular we decided to define two new points, $P_7$ and $P_6$, at which the system is  critical for temperatures $T=1/7$ and $T=1/6$ respectively, and then, with a linear interpolation, construct the corresponding quantity for the original point of phase transition $P$. 
To perform the simulations, we used a cluster-based nonlocal update algorithm, an adaptation of the Swendsen-Wang prescription, which is described in more detail in \cite{confanomaly}.

We used $L=200$ finite-temperature lattices to find the couplings corresponding to $P_6$ and $P_7$, and at such critical points we took $\mathcal{O}(10^5)$ measurements of the plaquette at 26 values of spatial side $L$, ranging from $L=10$ to $L=165$. 
The data fitted very well to the expectation from $L=70$ already, so we could extract two values of the critical index $\nu$:
\bea
	\nu_{T=1/6}  =  0.8004(19)[22]\nonumber \; , \ \ \ 
	\nu_{T=1/7} =  0.7942(18)[38] \; ,
\eea
in which the first uncertainty refers to the statistical fluctuations while the second is an estimate of the systematic error in the measurement. 

By linear interpolation along the couplings, the value of $\nu$ and the (coupling-dependent) critical temperature $T_c$ was calculated for the very point $P$. We found
\EQ
	\frac{T_c}{\sqrt{\sigma}} = 1.0393(12), \ \ \ \ \ \nu(P) = 0.7984(19)[27].
\EN

By plugging it into the formula for the mass ratio (\ref{eq:ratio_nu}), we obtain the following result for the mass ratio:
\EQ
	\frac{M_1}{M}(P) = 1.6124(71)[102] \; ,
\label{mfromnu}
\EN
which is well compatible with the less accurate estimate coming from the quantities in \cite{confanomaly} and thus well supports our conjecture.

\section{Conclusions}
In this paper we studied the ratio of the string tensions
$\sigma_2(T)/\sigma(T)$ near the deconfining point $T_c$ of a 3D
$\Z_4$ gauge model and compared the result with a general formula
which is expected to be true near $T=0$ for a generic gauge theory in
three or four dimensions. 
In this particular case we have combined numerical experiments with known 
exact results of  an integrable 2D quantum field theory that belongs, according to the Svetitsky-Yaffe conjecture, to the same universality class of the critical gauge system. 

An interesting property of the integrable model is that the mass ratio of the 
two physical states of the theory, which should equate the string tension 
ratio near $T_c$, can be expressed as a simple function of the thermal 
exponent $\nu$  (see Eq.(\ref{eq:ratio_nu})). 
We used two 
different methods to evaluate such a ratio, and both the estimates give compatible results which nicely agree with the ratio $\sigma_2/\sigma$ evaluated at $T=0$
(see Eq.(\ref{ratiosigma})). We then conclude that, at least in this model, 
the k-string tension ratios do not depend on $T$.

\end{document}